\documentclass[manuscript,screen]{acmart}
\AtBeginDocument{%
  \providecommand\BibTeX{{%
    \normalfont B\kern-0.5em{\scshape i\kern-0.25em b}\kern-0.8em\TeX}}}
\usepackage[commandnameprefix=always]{changes}
\setcopyright{acmlicensed}
\copyrightyear{2018}
\acmYear{2018}
\acmDOI{XXXXXXX.XXXXXXX}

\acmConference[Conference acronym 'XX]{Make sure to enter the correct
  conference title from your rights confirmation emai}{June 03--05,
  2018}{Woodstock, NY}
\acmBooktitle{Woodstock '18: ACM Symposium on Neural Gaze Detection,
 June 03--05, 2018, Woodstock, NY} 
\acmISBN{978-1-4503-XXXX-X/18/06}

\usepackage{changes}
\usepackage{float}
\usepackage{subcaption}
\begin{document}

\title{Beyond Functionality: Co-Designing Voice User Interfaces for Older Adults' Well-being}

\author{Xinhui Hu}
\authornote{Both authors contributed equally to this research and are co-first authors}
\email{xhu15@illinois.edu}
\orcid{0000-0001-9847-0788}
\affiliation{%
  \institution{University of Illinois at Urbana-Champaign}
  \city{Urbana-Champaign}
  \country{USA}
}

\author{Smit Desai}
\authornotemark[1]
\email{sm.desai@northeastern.edu}
\orcid{0000-0001-6983-8838}
\affiliation{%
  \institution{Northeastern University}
  \city{Boston}
  \country{USA}
}

\author{Morgan Lundy}
\email{melundy2@illinois.edu}
\orcid{0000-0002-7407-9378 }
\affiliation{%
  \institution{University of Illinois at Urbana-Champaign}
  \city{Urbana-Champaign}
  \country{USA}
}

\author{Jessie Chin}
\email{chin5@illinois.edu}
\orcid{0000-0002-2878-8544}
\affiliation{%
 \institution{University of Illinois at Urbana-Champaign}
  \city{Urbana-Champaign}
  \country{USA}
}

\renewcommand{\shortauthors}{Trovato and Tobin, et al.}

\begin{abstract}
The global population is rapidly aging, necessitating technologies that promote healthy aging. Voice User Interfaces (VUIs), leveraging natural language interaction, offer a promising solution for older adults due to their ease of use. However, current design practices often overemphasize functionality, neglecting older adults' complex aspirations, psychological well-being, and social connectedness. To address this gap, we conducted co-design sessions with 20 older adults employing an empathic design approach. Half of the participants interacted with a probe involving health information learning, while the others focused on a probe related to exercise. This method engaged participants in collaborative activities to uncover non-functional requirements early in the design process. Results indicate that when encouraged to share their needs within a social context, older adults revealed a range of sensory, aesthetic, hedonic, and social preferences and, more importantly, the specific personas of VUIs. These insights inform the relative importance of these factors in VUI design.

\end{abstract}




\maketitle

\section{Introduction}

Our world is experiencing a significant demographic shift, with the United Nations projecting that by 2060, individuals aged 65 and over will comprise a quarter of the global population \cite{Beard_Officer_Cassels_2016}. The human aging process brings various changes in both mind and body, such as decaying vision, declining mobility, and declined cognitive abilities \cite{kang2022ageism}. This aging trend has underscored the importance of leveraging technology to support healthy aging, with Voice User Interfaces (VUIs) emerging as a promising avenue for innovation and intervention \cite{sin2021vui, Kim_Choudhury_2021, desai2023ok, Pradhan_Lazar_2021}. VUIs are voice-based conversational agents widely implemented in smart devices such as smartphones, tablets, and smart speakers. Engaging users through auditory channels, VUIs often resemble the natural structure of daily conversation and thus present a minimal learning curve to users, positioning this technology as ideal for older adults who may experience age-related declines in visual acuity or motor skills \cite{Sayago_Neves_Cowan_2019}. Research has shown that older adults often express a preference for VUIs due to their ease of use compared to other modalities, and some even perceive these agents as fulfilling social roles akin to companions \cite{Pradhan_Findlater_Lazar_2019}. In recent years, substantial progress has been made in leveraging the potential of VUIs to facilitate informal learning, medication instruction, companionship, and skill development among older adults \cite{desai2023ok, Bickmore_Caruso_Clough-Gorr_Heeren_2005, Mathur_Dhodapkar_Zubatiy_Li_Jones_Mynatt_2022}.

Nevertheless, to ensure a seamless and enjoyable VUI experience for older adults, it is crucial to avoid the common design pitfall of overemphasizing functional limitations and constraints when addressing the needs of users with special requirements. This tendency often fosters the misconception that their lives are solely defined by these challenges rather than their broader aspirations, lifestyles, and personal preferences. As early as 2003, Coleman, Lebbon, and Myerson \cite{coleman2003design} highlighted this disparity, noting a significant gap between mainstream design's focus on the psychological and emotional factors motivating consumer choices and the emphasis on compensatory aids and adaptations for older and disabled individuals. Echoing this concern, Huppert \cite{huppert2003designing} also observed that design research often focuses on older adults in isolation, overlooking the rich social context in their lives that has profound implications for understanding their living situations and preferences. The risk with this tunnel vision on functionality is that designers may adopt a "frailty narrative" \cite{Sin_2022}, prioritizing concerns about impairment and compensatory functions even outside the medical context. Users with special needs are thus sometimes positioned as passive recipients of care-taking in design practice, with their agency and broader needs beyond functionality overlooked or relegated to a lower priority \cite{trevena2024vr}. Such a tendency inadvertently marginalizes users with special needs by treating them as separate from mainstream consumers, creating a sense of "otherness."

In seeking to move beyond this obstacle, the design domain is increasingly advocating for a paradigm shift when designing for older adults, moving away from a frailty narrative and moving towards embracing a more inclusive mental model. This approach encourages designers to view older users as active participants in their communities who possess rich social connections, varied interests, and aspirations for fulfilling experiences that extend far beyond basic functional needs \cite{coleman2003design, grundy2018supporting}. In illustrating their differences, Coleman \cite{coleman2003design} offers a compelling example: when designing a kitchen for a friend with physical challenges while appreciating all functional considerations, the friend's primary desire was to have a kitchen that would make the neighbors envious. This anecdote serves as a powerful reminder that users with special needs perceive themselves as ordinary consumers who place significant value on the non-functional, aesthetic, and social aspects of design that are shared regardless of age and capabilities. By embracing a more inclusive mental model, designers can create solutions that balance functional considerations and non-functional preferences and foster a sense of belonging, self-expression, and overall well-being for older adults, recognizing and celebrating their full humanity and agency.

While the development of VUI is not severely hindered by overlooking critical non-functional aspects, it stands to gain significantly from incorporating the social model of design. The current field of VUI design has already benefited from well-established guidelines and heuristics addressing key functional considerations, such as help and support, error recovery, context preservation, and output consistency \cite{Langevin_Lordon_Avrahami_Cowan_Hirsch_Hsieh_2021}. Notably, Murad et al. \cite{Murad_Candello_Munteanu_2023} consolidated published guidelines from  336 articles into 14 synthesized guidelines, including privacy, cognitive load, error rate, and usability. However, these guidelines largely overlooked hedonic factors and emotions, limiting the discussion of emotional factors to superficial elements such as encouragement and humor. Consequently, non-functional elements—particularly hedonic, aesthetic, and social needs—remain underexplored, allowing deeper integration to enhance user experience.

This opportunity aligns with recent VUI research indicating that older adults desire authentic two-way "human" conversation from VUIs, a goal transcending mere functional and usability requirements. A recurring theme in prior studies reveals that older adults seek sociability and even personality from VUIs \cite{Desai_Twidale_2023, li2021pueva, jeong2015s}. These personalities essentially serve as the front-end of VUIs and are referred to as 'system personas' or simply 'personas' \cite{Desai_Twidale_2023}. Importantly, this doesn't imply an expectation of advanced artificial intelligence with autonomy; rather, it reflects older adults' sensitivity to unnatural elements in the interaction and their tendency to disengage if they perceive the VUI as overly robotic. Understanding their aesthetic and social preferences is thus crucial for creating a natural and engaging conversational experience. Research has shown that factors such as the gender of a VUI's voice can influence users' perceptions of information reliability and accuracy \cite{jeong2015s}. Similarly, speech pace \cite{wang2023talk}, levels of politeness and assertiveness \cite{li2021pueva}, speech formality \cite{Chin_Desai_Lin_Mejia_2024}, and personas \cite{Desai_Dubiel_Leiva_2024} have all been identified as significant elements shaping the overall user experience with VUIs. Developing appropriate personas for VUIs is crucial, as it encompasses many of these non-functional aspects and can significantly impact user engagement and satisfaction \cite{Murad_Candello_Munteanu_2023}. 

However, unlike easily articulated functional needs, non-functional requirements are often latent and responsive, eluding traditional direct inquiries. To address this gap, we adopted a participatory co-design approach, engaging older adults in designing VUIs for their peers. This method creates a flexible environment for participants to explore and articulate their preferences while providing a structured framework to organize insights. Our approach emphasizes balancing functionality with \textit{Kansei} needs---a Japanese concept encompassing the emotional, sensory, and aesthetic aspects of product interaction. By integrating co-design \cite{ostrowski2021personal} with Kansei principles \cite{coleman2003design}, we aim to engage older adults as key stakeholders in technology development, ensuring their diverse needs are meaningfully captured and addressed in VUI design.

In this study, we implemented a two-phase approach. First, we introduced VUI-based design probes in health-related contexts given the increasing needs of the well-being of older adults: half of our 20 older adult participants engaged with a probe for learning health information, while the other half interacted with a physical activity guidance probe. These probes served as experiential primers, familiarizing participants with VUI applications in relevant scenarios before the co-design phase. This preparation aimed to broaden participants' understanding of VUI capabilities, providing baseline knowledge to fuel informed and innovative contributions. In the subsequent co-design session, all participants collaborated to explore VUI needs from a design perspective. This approach allowed us to explore the Kansei needs that might remain unexpressed, potentially revealing how different VUI exposures influence design ideas and perceived needs among older adults.

To this end, we aim to address the following research questions (RQs): \textbf{RQ1}, \textit{how do older adults and researchers perceive and collaborate in the co-design process in the context of health information via VUI?}; \textbf{RQ2}, \textit{what are the priorities of older adults regarding the presentation of health information, including content and involvement of VUI?}; and \textbf{RQ3}, \textit{what is an appropriate system persona in the context of designing agents for well-being?} These research questions are interconnected, with RQ1 focusing on the co-design process itself, RQ2 addressing the specific content and interaction preferences of older adults, and RQ3 exploring the crucial aspect of VUI persona development. Together, they provide a comprehensive approach to understanding and improving VUI design for older adults in health contexts.

Building on the findings from our RQs, our study offers three primary contributions to the landscape of VUI design and research, offering actionable insights and strategies for researchers, practitioners, and conversational designers:

\begin{itemize}
\item \textbf{Co-Design Process Insights for Researchers:}
We illuminate the dynamics of collaborative design with older adults, offering strategies for effective engagement in the co-design process. Our work reveals solutions to common challenges in facilitating design sessions with older adults and provides methodological considerations for future participatory studies with VUIs.
\item \textbf{Implementation Priorities for Practitioners:}
We provide actionable insights for those implementing VUIs for older adults' health and well-being needs. This includes key considerations for content presentation and system functionality, guidelines for balancing functional and non-functional requirements, and strategies for aligning VUI implementations with older adults' preferences and needs.
\item \textbf{System Personas for Designers:}
We present archetypes for conversation designers who are creating VUI personalities for older adults. This encompasses persona descriptions that resonate with older adults in health contexts, rationale for persona development based on user preferences, and guidance on integrating these personas into health-based VUI applications.
\end{itemize}

Collectively, these contributions bridge the gap between research, implementation, and design of VUIs for older adults' well-being needs. By addressing the entire spectrum from co-design processes to practical implementation and persona design, our work makes much-needed efforts to create more effective, user-centered voice interfaces in this critical domain.

\section{Related Work}
\subsection{Older adults with VUIs }
VUIs, such as Amazon Alexa, Apple Siri, and Google Assistant, are gaining traction among older adults as a user-friendly alternative to the complexities of Graphical User Interfaces (GUIs). This shift is particularly significant given the global aging trend, prompting the United Nations to declare 2021-2030 the decade of healthy aging \cite{Beard_Officer_Cassels_2016}. Research has consistently linked technology use by older adults to improved physical and mental health, increased autonomy, and reduced feelings of loneliness \cite{Cody_Dunn_Hoppin_Wendt_1999, Czaja_Lee_2006,Mihailidis_Carmichael_Boger_2004, Wang_Bolling_Mao_Reichstadt_Jeste_Kim_Nebeker_2019}. 

Despite these benefits, misconceptions about older adults' relationship with technology persist, suggesting that seniors are disinterested in modern technology, find it useless, lack the ability to use it, or are difficult to teach \cite{Wandke_Sengpiel_Sönksen_2012}. However, these misconceptions often stem from overgeneralized design and inadequate technological support rather than inherent limitations of older adults \cite{Czaja_Lee_2006,Heinz_Martin_Margrett_Yearns_Franke_Yang_Wong_Chang_2013,Wandke_Sengpiel_Sönksen_2012}. VUIs offer a more accessible modality compared to GUIs, particularly for individuals facing sensory and cognitive challenges, difficulties with traditional input devices, visual impairments, or lack of confidence in navigating complex interfaces \cite{Portet_Vacher_Golanski_Roux_Meillon_2013, Pradhan_Lazar_Findlater_2020}.

Older adults have demonstrated positive perceptions of VUIs, appreciating their companionship and conversational features and showing decreased concern about making mistakes over time \cite{Kim_Choudhury_2021,Pradhan_Findlater_Lazar_2019}. However, VUI adoption among older adults is not without challenges, including auditory issues, unfamiliarity with VUI conversational norms, conversational breakdowns, and privacy concerns \cite{Pradhan_Findlater_Lazar_2019,Pradhan_Lazar_Findlater_2020,Kim_Choudhury_2021}. The aging process brings physiological changes and social stigmas that can lead to feelings of alienation, anxiety, and depression \cite{Holm_Lyberg_Severinsson_2014}. As global demographics shift towards an older population, the role of technology in mitigating the impacts of aging and supporting aging in place becomes increasingly crucial. However, the literature reveals significant gaps in VUI design for older adults, including the need for novel applications to enrich their experiences, understanding long-term behavior with VUIs, and dispelling myths about their technological capabilities.

To address these needs, recent studies have adopted participatory design approaches, incorporating older adults into the VUI design process. These initiatives, including design workshops and co-design activities, aim to tailor VUI technology to the specific requirements of older users, enhancing usability and overall experience \cite{Desai_Lundy_Chin_2023, Harrington_Wilcox_Connelly_Rogers_Sanford_2018, Harrington_Borgos-Rodriguez_Piper_2019,Harrington_Garg_Woodward_Williams_2022}. This participatory approach is crucial for overcoming VUI adoption challenges among older adults and leveraging technology to support healthy aging and independent living. 

\subsubsection{Co-designing with older adults}
Co-designing is a participatory approach fostering collective creativity throughout the design process \cite{steen2013co}, manifesting in various forms with varying levels of older adult involvement \cite{sumner2021co}. It involves designers, end-users, experts, and other stakeholders, aiming to enhance creativity and gain valuable user insights. Engaging older adults in co-design ensures their needs are accurately represented, contrasting with traditional approaches that often relegate users to passive roles. Coleman et al. highlighted the challenge for young designers to understand the complex needs of people outside their peer groups \cite{coleman2003design}. This insight addresses the tendency to overcompensate for perceived problems, potentially leading to defining older adults solely by their limitations. In extreme cases, older adults were characterized in terms of "illness, dependency, and decay" or "positioned as users exhibiting reluctance and resistance to new technology," sometimes even framed as "technologically illiterate and ignorant" \cite{peine2014rise}. Co-design emerges as a solution by fostering equitable designer-user relationships.

Research has successfully integrated older adults into various co-design scenarios, from need assessment \cite{davidson2013participatory} and prototyping \cite{hakobyan2013designing} to product testing and evaluation \cite{iacono2014engaging}, validating this approach despite initial doubts about older adults' proficiency as co-designers for technology agents \cite{sumner2021co}. The role reversal empowers older adults to express preferences more freely \cite{sanders2008co} and helps challenge stereotypical assumptions \cite{sakaguchi2021co}. This is particularly relevant when working with older adults, as their involvement bridges knowledge gaps stemming from generational differences in experiences. Moreover, by encouraging users to approach their needs from a designer's perspective, co-design activities promote design thinking, fostering creativity and enhancing users' ability to articulate their needs more effectively \cite{adikari2016embed}. When designers and users come from different communities, co-design serves as a valuable opportunity to verify the identification of the correct design problem and challenge stereotypical assumptions, ultimately leading to more inclusive and effective design solutions.

Despite its benefits, the open-ended nature of design tasks can be daunting for older adults \cite{Harrington_Borgos-Rodriguez_Piper_2019}, particularly when working with unfamiliar technologies like VUIs. This unfamiliarity can potentially constrain their ability to envision and articulate innovative design possibilities. To address this challenge and facilitate more informed ideation, we incorporated innovative design probes into our methodology \cite{hutchinson2003technology}. These probes serve as experiential primers, offering participants concrete interactions with VUI applications in relevant contexts. By providing hands-on experience prior to the co-design phase, we aimed to expand participants' understanding of VUI capabilities and potential applications, thus broadening the spectrum of possibilities they could envision and discuss during the subsequent co-design sessions.

The negative effects of group dynamics in co-design processes have been consistently emphasized across studies \cite{sakaguchi2021co, iacono2014engaging}. Informed by these insights and the need for in-depth individual engagement, our study adopted a one-on-one co-design approach between participants and the experimenter. We were further inspired by Slattery et al. \cite{Slattery_Saeri_Bragge_2020}, who identified a lack of process descriptions from the co-designer-as-facilitator perspective in health-related co-design studies, despite the method's increasing popularity. Given the limited established best practices for co-designing with older adults in health contexts, we employed a reflection-in-action approach \cite{schon2017reflective}. This method, previously applied in mobile health application design studies \cite{wechsler2015hcd, tong2022lessons}, not only enhanced our practice but also generated valuable in-action observations and reflections.

These considerations and methodological choices form the foundation for \textbf{RQ1}, which explores how older adults and researchers perceive and collaborate in the co-design process within the context of health information via VUI.

\subsection{Approaching Non-Functional Needs}
\subsubsection{Probing the Non-functional Needs}

Recognizing the unmet non-functional needs of older adults is crucial in designing effective VUIs. An examination of the literature reveals several interconnected obstacles contributing to their disengagement from digital technologies, spanning societal perceptions and personal anxieties. Negative attitudes and stereotypes about aging \cite{kang2022ageism} often intersect with feelings of perceived isolation or helplessness \cite{chaudhuri2017older}, compounding the challenges faced by older adults. The loss of social contact and assistance \cite{heinz2013perceptions} further exacerbate these issues, while a lack of technological aesthetic appeal \cite{orellano2016hispanic} can make digital solutions less inviting. Many older adults experience digital alienation and fear of social disapproval \cite{kong2018smart}, intensified by a perceived lack of emotional reciprocity in digital interactions \cite{bajones2019results} and fear of losing valuable face-to-face interactions \cite{lindberg2021older}. These interconnected obstacles highlight critical areas where non-functional needs are not being adequately addressed, underscoring the complexity of designing digital technologies that truly resonate with older adults' holistic needs and preferences.

Internalized negative attitudes can lead to isolation and diminished self-worth \cite{kang2022ageism}, while technological frustrations may be perceived as personal failures \cite{kong2018smart, desai2023using}. This highlights a disparity in how technological challenges are perceived; while younger generations might view them as neutral technical glitches, older adults might interpret them as personal inadequacies. Current products often create a perceived trade-off between technological interaction and interpersonal communication \cite{lindberg2021older, huppert2003designing}, reflecting a longstanding tendency in design research to focus on older individuals in isolation. This issue is further compounded by the challenges older adults face when adopting new technologies, including steep learning curves, anxieties about their ability to interact effectively, and the negative impact of lacking social support during the learning process \cite{kebede2022digital}. Addressing social needs, therefore, goes beyond merely connecting older adults to others through technology; it requires understanding the broader ecological context that fosters meaningful engagement, including their existing social networks, physical environments, and the emotional and psychological factors influencing their technology interactions.

Designs targeting older adults often fall short in terms of aesthetics and hedonic value \cite{orellano2016hispanic}, overemphasizing functionality inadvertently relegating enjoyment to a secondary role. This can result in designs prioritizing practicality over pleasure, potentially limiting the overall user experience. To holistically address these diverse non-functional needs in VUI design, it's critical to ensure older adults feel valued and empowered, recognized as active and discerning consumers whose preferences matter, rather than passive recipients of care or technology \cite{trevena2024vr}. Kansei design, with its focus on capturing and translating users' feelings toward a product, emerges as a promising strategy to address these complex needs, potentially bridging the gap between functionality and the deeper, more nuanced experiential aspects of technology interaction for older adults.

\subsubsection{Meeting the Needs with Kansei Design}

Kansei design is an approach that translates users' sensory and emotional experiences into design decisions. Originating from the Japanese concept of "emotion" or "sensibility," Kansei encompasses intuition, feelings, aesthetics, and sensitivity \cite{schutte2023Kansei}. The approach gained prominence in 1986 when Mazda branded a new car model based on engine sounds rather than purely functional advancements. In Kansei design, designers integrate users' subjective feelings, preferences, and emotional responses into design elements, aiming to create products that evoke specific emotional or aesthetic experiences \cite{lee2002pleasure}. Unlike purely functional or sensory design, Kansei design goes beyond ergonomic considerations to address aesthetic, emotional, psychological, and hedonistic user needs \cite{lee2002pleasure}. For example, in designing a smart speaker with a 'squircle' shape, a functional approach might focus on easy gripping and injury prevention, while a Kansei perspective would consider users' aesthetic satisfaction and the social value of showcasing the device.

Functionality and Kansei design are complementary rather than mutually exclusive. Designers must balance "objective and subjective properties, functional technology and emotional expressiveness, information and inspiration" \cite[p.1]{lee2002pleasure}. Typically, functional design precedes Kansei considerations, with the latter becoming more influential once basic functionality is achieved. VUI design is currently transitioning into this stage, as recent studies show older adults seeking sociability and personality from VUIs, disengaging when interactions feel synthetic \cite{desai2023using, li2021pueva}. 

These insights into the non-functional needs of older adults and the potential of Kansei design to address them highlight a critical area for investigation in VUI design. While we understand the importance of these needs, there remains a gap in our knowledge about how they specifically manifest in the context of health information delivery via VUIs. This led us to formulate \textbf{RQ2}, which explores the priorities of older adults regarding the presentation of health information, including content and involvement of VUI.

\subsection{System personas in VUIs}

In the field of conversational agents, certain design elements often evoke well-known characters from the annals of popular culture. Examples include the notorious HAL9000, the efficient "Computer" from Star Trek, or the ominous Big Brother. Such parallels, as identified by scholars \cite{azevedo2018using, Axtell_Munteanu_2021, Samrose_Anbarasu_Joshi_Mishra_2020}, are typically unintended by designers but are instead the result of the agents' inherent features that users connect to these fictional icons. These connections are strategically utilized by designers to guide user interactions towards positive outcomes, such as encouraging the active engagement seen with Star Trek's "Computer," while also addressing and alleviating negative connotations, such as the privacy concerns reminiscent of Big Brother.

Expanding on this theme, Google's Conversational UX guidelines\footnote{https://web.archive.org/web/20240601194224/https://developers.google.com/assistant/conversation-design/create-a-persona} propose the creation of system personas through 'role production.' This approach involves crafting a conversational agent that takes on a specific societal role, like an Exercise Coach or a Bank Teller, thus avoiding direct mimicry of any cultural icons. Modern voice-based agents, for instance, are often positioned as virtual "assistants" to foster a sense of utility and service. The choice of metaphors, such as "assistant," "servant," or "butler," however, can sometimes be problematic. They may inadvertently reflect and perpetuate historical biases linked to gender or race, as discussed in recent studies \cite{McMillan_Jaber_2021,Desai_Twidale_2022,Desai_Twidale_2023,Pradhan_Lazar_2021}. The use of system personas goes beyond mere role assignment; it taps into the deeper human tendency to attribute human-like qualities to non-human entities—a concept firmly rooted in the anthropomorphic literature within Human-Computer Interaction (HCI).

 Crafting an appropriate persona is a crucial aspect of conversation design, particularly in VUI development. Sadek et al. \cite{Sadek_Calvo_Mougenot_2023} revealed that VUI designers strongly prefer having access to predefined 'archetypes' or 'personalities' to facilitate the design process. This typically involves selecting descriptors, metaphors, biographies, and attributes like voice, age, and gender, sometimes even drawing inspiration from popular culture characters. Recognizing the importance of persona development, our study focuses on gathering and analyzing the perspectives of older adults to provide conversation designers with a rich, contextually relevant foundation for creating personas in the health domain. 
 
 By exploring older adults' preferences and expectations for VUI personalities, we aim to offer insights that resonate with this demographic, potentially enhancing the relevance and effectiveness of VUIs in their daily lives. These considerations drive our third research question, \textbf{RQ3}, which investigates the characteristics of an appropriate system persona in the context of designing agents for well-being, specifically tailored to the needs and preferences of older adults.

\section{Method}
\subsection{Participants}
We recruited 20 participants from a midwestern city in the United States for a 90-minute study, utilizing university mailing lists and newsletters. The participants comprised a diverse group, including 13 females, 5 males, and 2 individuals preferring not to disclose their gender, with ages ranging from 63 to 81 years old. The average age was approximately 69.05. All participants were reimbursed with \$20 Amazon gift cards for their participation. The majority were native English speakers and possessed a college degree. In terms of ethnic and racial backgrounds, most identified as white or Caucasian, with one participant identifying as both White or Caucasian and Native American or Alaska Native, and a few chose not to disclose this information. Familiarity and usage of VUIs varied, with most participants (85\%) having little to no prior experience, while a few used VUIs regularly (10\%), including daily users (5\%). The overall expertise with VUIs was low to moderate, with exceptions for a few participants who considered themselves more proficient (20\%).

\subsection{Apparatus and Materials}
\subsubsection{Apparatus}

\begin{figure}[h!]
  \centering
  \includegraphics[width=0.5\textwidth]{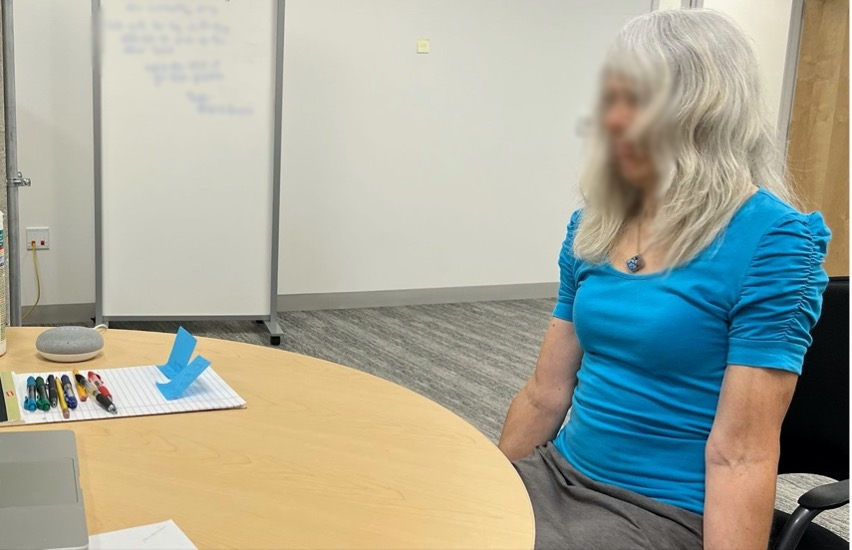}
  \includegraphics[width=0.3\textwidth]{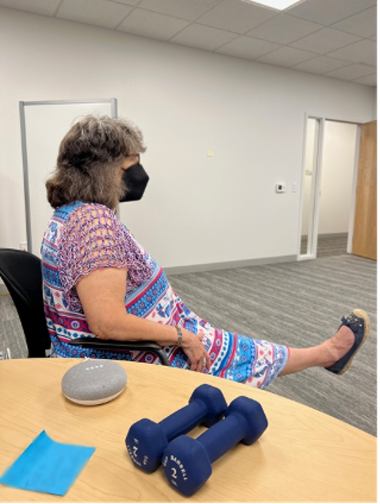}
  \caption{Pictures of participants seated in front of a Google Mini Smart Speaker in the lab setting, with various pens, papers, whiteboards, markers, and weights provided to them.}
  \label{fig1}
\end{figure}

In our study, we employed the Google Nest Mini Gen 2 smart speaker as our primary device, given its targeted marketing towards older adults and its widespread adoption within this demographic \cite{sin2021vui}. The studies were conducted in a research lab, with the smart speaker placed on a table in front of the participants to facilitate interaction. To accommodate participants' comfort and preferences, seating options varied between comfortable chairs with or without arms. The setup was further supplemented with materials designed to engage participants in the studies' activities. These included pens, papers, markers, and a whiteboard for note-taking and co-design activities, as well as handheld dumbbells, notebooks, sticky notes, and dry-erase markers to provide participants with options for physical engagement during the user study. The sessions were audio-recorded on Zoom with participants' consent, ensuring their interactions and feedback were recorded for further analysis. The setup of the lab can be seen in Figure \ref{fig1}.

\subsubsection{Demographic Questionnaire} 

In our study, we employed demographic questionnaires to gather data about our participants' backgrounds, including age, gender, ethnicity, occupation, English proficiency, and their familiarity and prior experiences with commercial VUIs like Amazon Alexa or Apple Siri.  

\subsubsection{Design Probes}
We used initial interactions with two different VUI prototypes as design probes to inspire ideas in the following co-design process. One VUI was designed to provide participants with an auditory health story with frequent pauses for question-and-answer interactions to promote playful interactive learning of health information. The second VUI was designed to provide participants with auditory directions to follow each step of multiple exercises, also with frequent conversational and check-in opportunities, to promote easy and enjoyable home exercise. Hereafter, we describe the two VUIs as 'learning probe' and 'exercise probe.'

\subsection{Procedure}
In our research, we adhered strictly to the guidelines set forth by the Institutional Review Board. Participants were initially briefed about the study's procedures, including the co-design activities, ensuring they understood their right to withdraw at any point. Following this briefing, participants electronically signed informed consent forms and completed demographic questionnaires. Before engaging with the VUIs, participants had the opportunity to familiarize themselves with basic commands using the Google Nest Mini Gen. 2, such as requesting weather updates. This initial interaction aimed to make participants comfortable with the technology and the conversational nature of the VUIs, setting the stage for more complex interactions.

\begin{figure}[H]
    \centering
    \includegraphics[width=\textwidth]{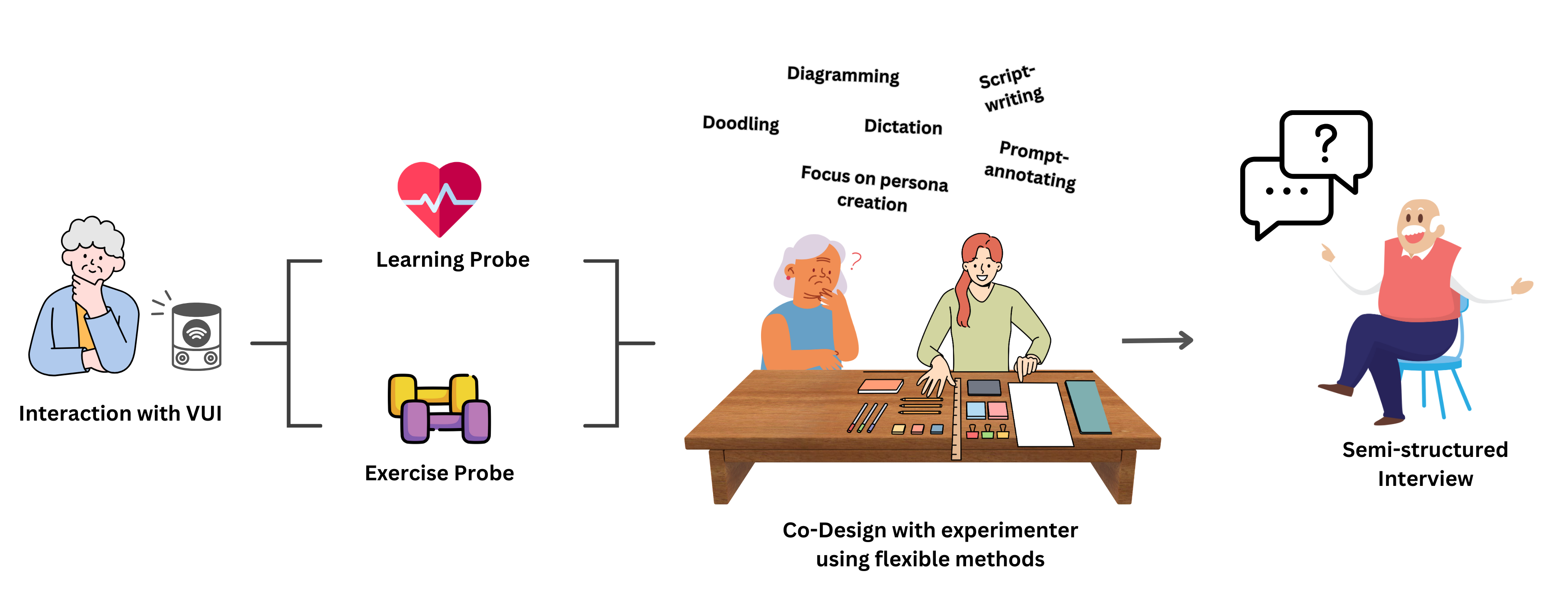}
    \caption{The Experiment Procedure Flow}
    \label{fig_flow}
\end{figure}

We used initial interactions with a VUI prototype as design probes to inspire participant ideas in a co-design process, as shown in Figure \ref{fig_flow}. All participants began the session by interacting directly with the prototype for 10-15 minutes. Half of the participants interacted with the health learning prototype, the 'learning probe,' where they navigated a health story, answered health-related questions, and used features like stop, replay, and menu. The other half of the participants interacted with a home exercise prototype, the 'exercise probe,' where they selected a home exercise to follow along with the VUI physically, and similarly used question-and-answer, stop, replay, and menu features. These two types of VUIs provided participants with a starting point for features they liked and disliked, a clearer understanding of what is achievable with VUIs, and ideas to bring to the next co-design session. Having two different prototype interaction probes prompted wider ideas across potential well-being applications of VUIs. 

All participants then discussed their thoughts on their assigned prototype experience with the researcher for 10-15 minutes, which both clarified the participants' own priorities and ideas and gave the researcher more information to better facilitate the following 30-minute co-design session. Questions included assessing what they liked and disliked about the experience, but also further leading questions about what they would prefer to replace or extend characteristics. Participants also reflected on and imagined the potential of their VUI design in their everyday lives.

\subsubsection{30 minute Co-design Sessions} 
The co-design session, lasting approximately 30 minutes, placed participants in the role of designers, aiming to understand their priorities and preferences for health-related contexts. Participants were tasked with creating or imagining dialogues between their ideal VUI and a person interested in either learning health information or guidance through an exercise.

Participants were handed a physical sheet of paper with the printed co-design prompt at the top and given a few minutes to read the prompt. The participants with the 'exercise probe' were given a prompt with step-by-step instructions for an overhead arm raise exercise, with steps like "keep your feet flat on the floor shoulder's width apart," how to hold the weights and for how long, and when to breathe and rest. The prompt also included a statement on why the exercise is beneficial. The 'learning probe' co-design prompt was more text-intensive, with a paragraph of complex information about Alzheimer's disease in a  style. The prompt began with the explanation that "your task is to write a scene where a person is talking to an expert voice AI to learn about the disease."
 
\begin{figure}[H]
    \centering
    \begin{subfigure}{0.45\textwidth}
        \centering
        \includegraphics[width=\textwidth]{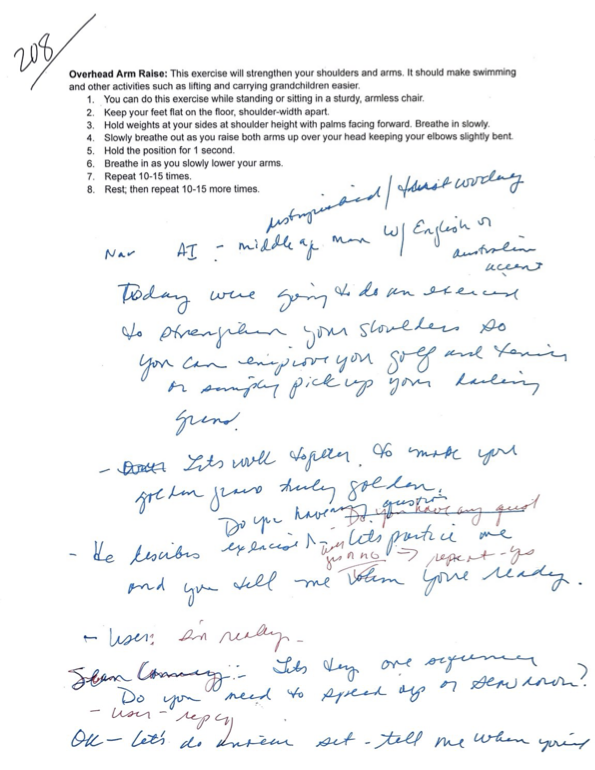}
        \caption{Co-designers composed scripts using multicolored pens to imagine the statements needed to create their ideal VUI characteristics.}
        \label{fig2}
    \end{subfigure}
    \hfill
    \begin{subfigure}{0.45\textwidth}
        \centering
        \includegraphics[width=\textwidth]{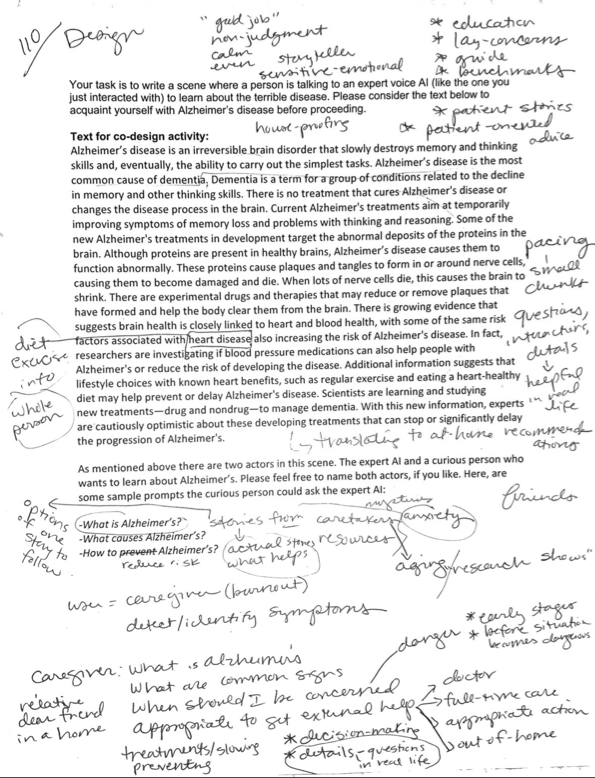}
        \caption{Co-designers annotated prompts to imagine user and health expert AI conversations, and brainstormed corresponding VUI characteristics.}
        \label{fig3}
    \end{subfigure}
    \caption{Co-designers taking two different approaches to creating co-design products while imagining user and AI conversations and creating VA characteristics, using the co-design prompts and materials provided.}
    \label{fig:combined}
\end{figure}

Co-design tasks were supported by various materials, such as colorful pens, papers, and markers, to aid participants in articulating their design concepts. Participants could annotate the prompt itself, as in Figure \ref{fig3}, or write out a full script between VUI and an imagined user beneath the prompt, as in Figure \ref{fig2}. Participants could also write on new sheets of lined paper, doodle VUI personas, diagram branching VUI characteristics, or write on a whiteboard for ease of erasure. Additionally, participants could dictate their ideas directly to the researcher, ensuring that all participants, regardless of their comfort with writing or drawing, could contribute meaningfully to the co-design process.

At the end of the co-design session, participants were asked to reflect on and discuss their experience co-designing with the researcher, including what was enjoyable, difficult, or unclear. The final debrief questions further investigated what participants would design with infinite time and resources, and the participants were asked to reflect on any changes in their understanding of VUI after the co-design experience. 

\subsubsection{Reflection in action} 
To address potential assumptions and pitfalls in our co-design approach, we adopt a "reflection in action" approach to co-design \cite{schon2017reflective}. Reflective approaches to the co-design process have been employed effectively by designers in the health domain, such as post-collaboration reflections \cite{wechsler2015hcd} and analysis of field notes and recorded meetings with developers \cite{tong2022lessons} in the design of mobile health applications. During and after the co-design sessions with each older adult, field notes were taken about effective and ineffective strategies, observations on co-designers comfort, understanding, and engagement with the co-design tasks, and ideas for improved facilitation in the following sessions. From the participant side, at the end of the co-design session, participants were specifically asked to reflect on and describe their perceptions of the co-design experience. As co-design with older adults in health contexts is not a widely used approach, with little established information on best practices of co-design methods in general \cite{Slattery_Saeri_Bragge_2020}, a reflective approach both improved our own practice and generated in-action observations and reflections that we hope will aid future researchers. 

\subsubsection{Researcher positionality statement} 
The researcher conducting the co-design sessions is a 30-year-old white and queer researcher in the U.S. As a disabled person and scholar, her approach to research is particularly defined by approaches grounded in critical disability theory, including a commitment to participatory methods that ensure 'nothing about' social groups is determined by researchers 'without' the inclusion of the relevant groups in the research process, an approach to HCI discussed in \cite{spiel2020nothing}. We therefore built our co-design approach to this study from values of the inclusion of those most affected by design in the design process \cite{chivukula2020bardzell}, in this case, older adults. A reflection-in-action approach \cite{schon2017reflective} prompted self-introspection on the researcher's part to identify any issues, knowledge gaps, or assumptions born of her positionality as they emerged during and after each co-design session.

Particularly relevant for this study, as a younger researcher with an invisible disability, effort was dedicated to making sure older adult co-designers felt valued, respected, and like true collaborators to preempt design paralysis, specifically in reaction to several older adult participants' apologies for their perceived negative relationship with technology in early sessions. Our approach, informed by existing literature and impacted by the researcher's personal experiences as a disabled, relatively young researcher, emphasized the researcher taking the role of facilitator rather than expert, understanding our participants as collaborators and elders with expertise rather than research subjects \cite{costanza2020design}, and aspiring for a humility-based co-design approach emphasizing truly shared agency \cite{spiel2020thinking}. How these values were operationalized and specific recommendations are discussed in 4.1. 

\subsection{Data Analysis}
In our research, we conducted a qualitative analysis of the co-design sessions, including audio recordings recorded with participants' consent and products co-generated during co-designing. These recordings were transcribed and analyzed by teams of researchers experienced in qualitative methods and with backgrounds in HCI. To ensure rigor and reliability in our thematic analysis, we adhered to established guidelines, incorporating both inductive and deductive coding strategies to develop a comprehensive codebook. Field and reflective notes and images of co-design products were also analyzed for emergent themes.

Initially, researchers individually familiarized themselves with the data through reflexive journaling, a process spanning several weeks to engage with the participants' perspectives deeply. Initial codes were generated independently, focusing on identifying patterns and themes relevant to the co-design process. These initial codes were then discussed extensively in multiple data analysis sessions. This collaborative approach, underpinned by a negotiated agreement method, ensured that all discrepancies in code interpretation were resolved through consensus, enhancing the analysis's trustworthiness.

The final phase involved a detailed review by a third researcher, which led to the refinement of codes and the development of overarching themes and subthemes. Our findings are presented with a clear delineation of theme prevalence among participants, using descriptive terms such as "a few" (up to 20\%), "some" (21-50\%), "most" (51-80\%), and "nearly all" (more than 80\%) to indicate the proportion of participants reflecting each theme. This approach ensures transparency and provides a clear sense of the themes' significance and resonance among the study participants.

\section{Findings}
\textbf{RQ1: How do older adults and researchers perceive and collaborate in the co-design process in the context of health information via VUI?}

\subsection{Recommendations and reflections on engaging older adults in co-design} 

\subsubsection{Conversational and Question-and-Answer approaches to co-design}

A conversational approach to the co-design process created opportunities for participants to direct the process, verbalize the social and sensory characteristics of their ideal VUI persona and demeanor, and respond to probing questions to prompt deeper design ideation. Among the 20 participants, 13 explicitly communicated their inclination towards formulating and sharing design ideas via question-and-answer participation or by responding to requests for suggestions rather than engaging in other types of cooperation. Though the term 'design paralysis' wasn't explicitly mentioned, participants appreciated the conversational guidance provided by designers using questions and helped co-design participants externalize their thoughts progressively, thereby avoiding situations where uncertainty about where to begin might arise. The low-stakes back-and-forth setting allowed co-designers to share their more subjective needs and preferences, such as aesthetic and emotional considerations, that went beyond a written script. 

Another strategy when co-designers stalled or were unsure of what to consider next was to revisit their positive and negative experiences with the prototypes, to emulate or sharply redesign their ideal VUIs' personality, voice design, and narrative structure. The researcher providing dictation during the co-design process, if participants desired, was a specific strategy within the conversational co-design approach that allowed older adult co-designers to focus on verbalizing their ideas, enabling time and space for exploring participants' multifaceted preferences, through an appropriately spoken medium similar to the auditory nature of VUI. 16 of 20 participants took advantage of varying levels of dictation by the researcher. Co-designers also found the ability to 'sound out' ideas and concerns and conveniently communicate with designers to be beneficial [P1].
 
\subsubsection{Sensory-rich material opportunities in co-design}
In addition to the conversational approach, strategies to spur participation included giving co-designers a moment for ideation on their own before beginning discussion, surrounding co-designers with colorful and aesthetically pleasing sketching paper, lined paper, the printed prompt, markers, a whiteboard, sticky notes, colored pens, pencils, and the device all within reach, and providing instructions that participants can take multiple approaches to verbalizing, writing, or diagramming their design ideas. 5 co-designers wrote out scripts for interactions between imagined users and their expert VUI (e.g., Figure \ref{fig2}), diagrammed the prompt text itself (e.g., Figure \ref{fig3}), 4 drew branches for multiple question progressions, and 1 participant drew an illustration of the VUI. VUI hedonistic and aesthetic preferences were generally verbalized, but participants often listed individualized characteristics as marginalia.  

\subsubsection{Real world references and big ideas for ideal characteristics}
Another tendency that appeared in co-design activities is the preference for real-world references. A total of 10 participants conveyed that incorporating real-world circumstances and human experiences played a crucial role in anchoring and articulating their design ideas in the co-design process. More specifically, when the designer proposed concrete real-world references, such as a specific home-based scenario or in specific information-seeking circumstances, the participants were able to better explain their designs and associate the ideas with their needs.

Beyond the everyday, a strategy employed to create opportunities for design thinking was to ask co-designers what they would design if they had unlimited time and money outside of the 30-minute co-design session. These questions created space for discussion of above-and-beyond aesthetic, subjective considerations at high levels. For example, having the agent seem 'more human' and like 'it is listening' [P4, P6, P11, P12, P14, P18, P20], and to create a sense of 'relationship' were all major goals for participants [P12, P14, P17]. 

\subsubsection{Strategies for humility, comfort, and care in co-design}
Last, strategies for co-design were most successful when they created comfortable space for co-designers to imagine and design. This included creating a welcoming, material-rich environment, and “making space” \cite{spiel2020thinking} for co-designers’ ideas through humility in the question-and-answer format, by restating co-designers’ ideas, asking clarifying questions to check the researcher’s understanding, and avoiding imposing researcher interpretations or priorities. Care was also embodied through active listening, open body language, clear interest and affirmation in writing down ideas, and taking note of co-designers’ hesitancies, anxieties, and indications of stress. These care-based strategies, and at times careful personal disclosure by the researcher, allowed co-designers to more comfortably share design ideas from their own personal lived experiences while dispelling discomfort around co-designing with or feeling ‘tested by’ a younger, seemingly able-bodied researcher. For other co-designers, the most effective strategy to encourage comfort in design thinking was to ask questions about how an imagined \textit{other} older adult user might want to interact with their VUI, rather than themselves.

\subsection{Older adults perception of co-design} 
 Of the 20 co-designers, 14 viewed the co-design experience as transformative and educational in that the experience changed their overall perceptions and knowledge about VUIs. Co-design methods allowed participants space to explore new VUI applications [P2, P4, P5, P8, P13, P18], and imagine and individualize their ideal VUI's social and aesthetic characteristics. Five participants also explicitly discussed how the co-design experience is useful for designing tools that older adults actually want to use, which incorporate their aesthetic values and preferences [P2, P4, P5, P13, P17].

In addition to the method's generativeness in identifying co-designers' hedonistic expectations and needs in the VUI context, the co-design process itself was also perceived to be pleasurable and sensory-rich. Twelve of the co-designers also specifically described the co-design experience as fun, with one co-designer describing the co-design session as their “first chance to play” and explore new VUI possibilities [P3]. All participants were interested in participating in another co-design session in the future. 

\noindent
\newline
\textbf{RQ2: What are the priorities of older adults regarding the presentation of health information, including content and involvement of VUI?}

\subsection{Content}

Regarding how VUI should present health information to users, participants' ideas converged into two main streams: directly delivering credible facts and crafting stories that implicitly embedded the knowledge. Interestingly, these two streams are not mutually exclusive, as multiple participants envisioned that VUI could provide both based on specific conditions, as shown in Figure \ref{fig_content}

\subsubsection{Facts and Knowledge}

Participants in both the learning probe and exercise probe conditions showed a strong expectation for VUIs to deliver facts, knowledge, and research advancements, albeit with different emphases. For those who designed VUI for exercise, the sharing of facts and knowledge was mainly perceived as a strategy to motivate users to engage actively in health-related activities. Therefore, in their designs, these participants highlighted the practical value of information, particularly how to implement health-related activities [P14, P17] and the benefits they will bring [P11, P12, P15]. 

\begin{quote}
    “I would like the VUI to tell me what exercise I need, what benefits it can bring, and what part of my body will be involved in the process. This helps me to build a healthy habit.” [P15]
\end{quote}

Further, participants who interacted with the learning probe designed VUIs to deliver information about the risk factors [P6, P8], modifiable factors [P8], potential treatments [P3, P10], preventive measures [P10], and latest research advancement [P1, P3, P8] for specific syndromes. In rationalizing such design, multiple participants mentioned that these facts and knowledge can alleviate worries when encountering medical concerns. For example, one participant explained that VUI could enable users to prepare themselves before engaging with the medical establishment, thereby mitigating mental burdens brought about by uncertainties.

\begin{quote}
    “I will make my VUI tell users what will be the symptoms, what are the risk factors, and what should be prepared before you jump into the medical appointments…So I feel I am well-informed before I go to the doctor. I am not ignorant. I won’t easily get misled or confused.” [P6]

\end{quote}

Moreover, participants in this condition generally conveyed their expectation that VUI would provide source information alongside the knowledge due to the emphasis on its accuracy and credibility.

\begin{quote}
    “It will be nice if the VUI can provide the source of the information along with the knowledge and facts. Thus, I can rest assured that information is coming from an authoritative and trustworthy source. ” [P3]
\end{quote}

Moreover, participants in both conditions acknowledged the potential risks of overwhelming users. As a countermeasure, they proposed the idea of VUI customizing its content based on the user's comprehension level and stress when delivering health information. For instance, P6 and P8 both designed their VUIs to monitor the user's level of stress or comprehension and tailor the information accordingly. They suggested that VUI could share basic information when users feel stressed or uninterested, and deliver detailed knowledge when users are attentive, with the goal of avoiding potential feelings of being overwhelmed. In the same vein, P10 suggested that VUI should allow users to skip knowledge and facts when not immediately needed, stating, "Not all details are necessary. Some people will think, ‘I don’t care about protein in the brain; just tell me what to do’" [P10]. Additionally, P4 and P10 discussed the importance of considering indirect users, such as users' family members or caregivers, in the information-sharing conversation. For instance, P10 described a scenario involving a potential user with Alzheimer's, suggesting that the VUI could provide information to family members and caregivers to enhance the home environment to suit the ensued concerns.

\subsubsection{Story}

Apart from providing users with factual information, eight of ten participants who interacted with the learning probe also recommended incorporating a storytelling method. They suggested that the VUI can present health information by skillfully incorporating critical details within empathic narratives. This approach aims to facilitate easy and enjoyable conversations, especially when users are seeking to acquire medical knowledge in non-urgent situations. As one participant explained, "I hope the VUI will show something like an actual story, and not just like facts that you are trying to teach somebody. Knitting the facts into the story. That is very useful." [P5]

In the narratives created by participants, a recurring pattern emerged: symptoms emerging in an older family member have been noticed, leading the family to become aware of the associated syndromes and prompting them to seek medical treatment or intervention. Notably, when creating these informative stories, participants ideated from a remarkably varied array of third-person perspectives, which spanned from children [P4, P7] to grandchildren [P5], caregivers [P10], and medical professionals closely acquainted with the patient [P9].

\begin{quote}
    “I think a story of a mother will be powerful. A little history about her will be great, such as she was a nurse during World War II and she was a good swimmer with heart disease. Then, the story should shift to her recent situations, syndromes start to show and her kids take care of her…So after hearing this story, the users should be able to answer such a question: what should I do if I or any people I care about show the same syndromes.” [P4]

\end{quote}

\begin{quote}
      “The story can start with the grandchildren learning about their grandparent's syndromes from their parents, preparing them for the upcoming weekend visit. As the parents told the kids about the symptoms, the users of VUI will acquire the relevant knowledge.” [P5]
\end{quote}

\begin{quote}
        
    “I would like the story to start with a doctor attentively listening to a visitor discussing the concerned disease and providing medical advice, along with general recommendations. It will be ideal if the doctor has both the professional knowledge and knows the visitor personally so medical insights and daily-life scenarios will be covered.” [P9] 
\end{quote}

The rest of the participants acknowledged the cognitive impact of stories in aiding contextualization, comprehension, and memorization without diving much into the storyline. For instance, a participant highlighted that a story enables users to relax their minds, alleviating the need for constant concentration that is required when listening to a list of facts: "Story is like the water running under the bridge. If you make a small slip, it will not necessarily mess up your understanding of the story." [P6]

\subsection{Interaction Mode}
\begin{figure}[H]
    \centering
    \includegraphics[width=\textwidth]{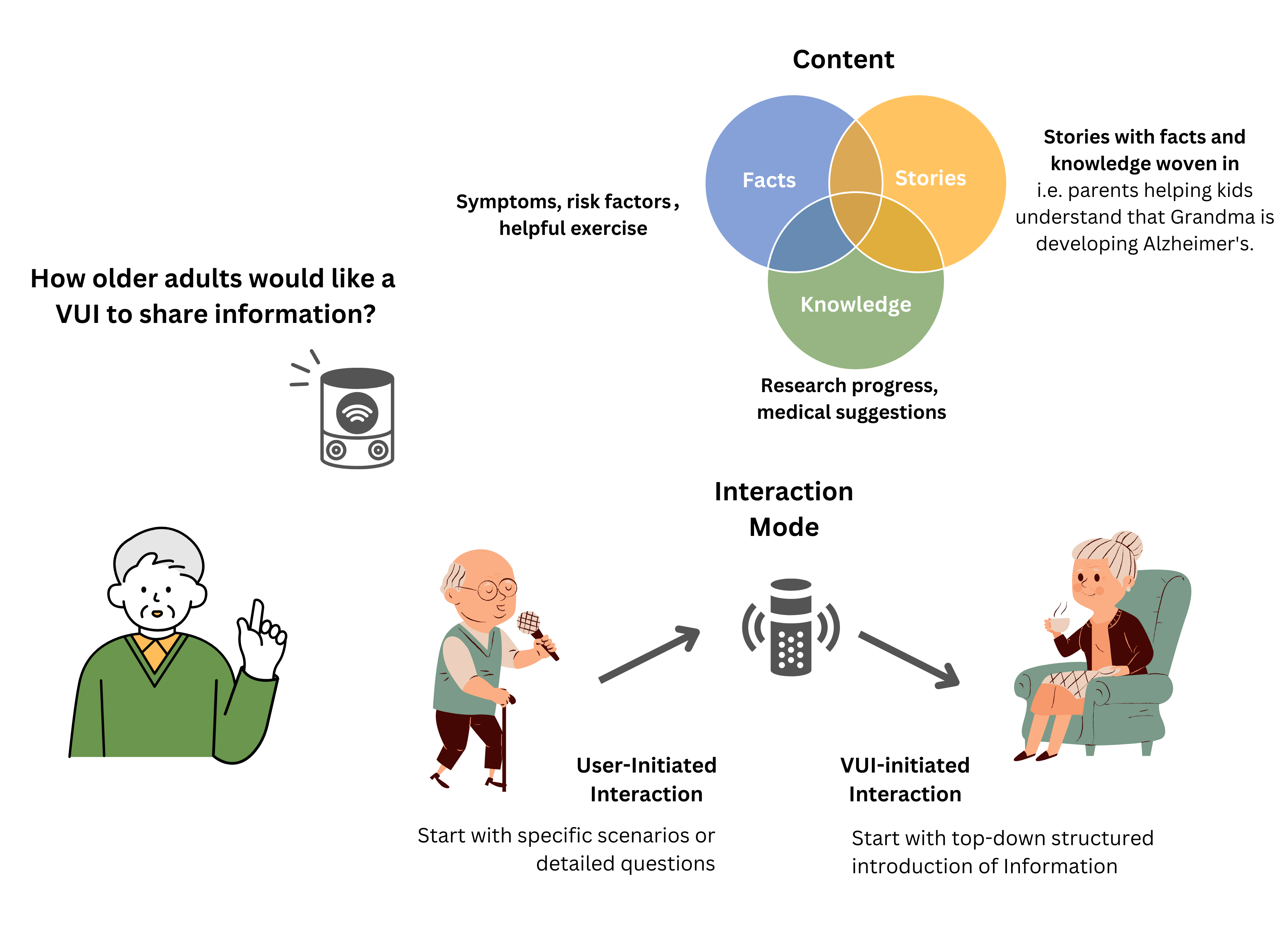}
    \caption{Older Adult Participants's Preferred Content and Involvement with VUI}
    \label{fig_content}
\end{figure}

When considering whether the VUI or the user should take an active role in navigating the conversation, the participants' designs also exhibited notable divergence, as illustrated in Figure \ref{fig_content}. Most participants in the exercise and some in the learning condition opted for the VUI to actively initiate and prompt conversations. The rationale behind this preference was grounded in the recognition that users might not always be aware of crucial questions, potentially leading to feelings of being overwhelmed about what to inquire. 

\begin{quote}
    “I don’t know if I would want to be that person who asks the question to VUI and get the answer back. Because I may not know the question to ask, so I would probably let VUI tell me.” [P9] 
\end{quote}

\begin{quote}
    “It would be nice if VUI could remind me me to begin my health activities for the day and ask whether my previous exercise made me feel better.” [P12] 
\end{quote}

On the other hand, participants favored users to take an active role in interaction and justified their preference for several reasons. For instance, one participant [P1] argued that users' questions tend to be less structured and cover a broader range than designers might anticipate, making it challenging to design the VUI to effectively initiate the conversation. Expressing a similar concern, one more participant emphasized that the questions initiated by the VUI might not always align with the users' most pertinent concerns. This participant preferred a scenario where users take the lead in questioning the VUI, suggesting queries like, "I am observing such and such phenomena, is this something I should care about? Or is it just normal?" [P7]  Furthermore, building on the concept that "the best way to learn is to teach,"  another participant suggested that users should have the chance to question the VUI about the validity of a medical belief instead of “just picking out a random or pre-programmed function.” [P6] For instance, this participant envisioned a scenario where a user might ask, "I think there is a saying about the heart, is it true?" The VUI, in turn, evaluates the questions and provides feedback, reaffirming the user's knowledge and correcting the mistakes.

\noindent
\newline
\newpage
\noindent
\textbf{RQ3: What is an appropriate system persona in the context of designing agents for well-being?}
\subsection{VUI Persona}
We adhered to Google's Conversation Design Guidelines in our persona creation process, ensuring alignment with industry standards while meeting our goal of providing practical archetypes for conversation designers. This approach bridges academic research with industry implementation, creating personas that reflect older adults' needs and integrate seamlessly into the conversation design workflow.
\subsubsection{Character}

In the co-design of the desired VUI, 14 out of 20 participants placed an explicit focus on ideating the personality and voice features. Participants’ consideration for the personality of VUI mainly revolved around the social roles it embodied. In brief, most of our participants assigned a professional, expert, or physical educator persona to the VUI in the co-design and showed converging emphases on how the professional persona of VUI will add trustworthiness and credibility to the information it shared in a health information sharing context. Out of the 20 participants, 8 believed that the VUI should be assigned professional roles for information seeking or health support, such as medical experts like doctors [P1, P4, P6], nurses [P8], or physical educators [P12, P15, P17, P18, P20]. The participants' rationale for this design preference is grounded in their aspiration to craft a knowledgeable and trustworthy personality for the VUI. This approach aims to make the VUI sound affirmative and persuasive when delivering information.

\begin{quote}
    “Definitely an expert persona, a doctor persona. I would like my information to come from someone who knows about their areas and topics.” [P1]
\end{quote}

On the other hand, outside the information-sharing scenarios, there are also several participants who favored a friend or companion persona [P9, P14, P20]. 

\begin{quote}
    “I would want them to be relaxed and somewhat friendly. Not totally professional like give me the facts and nothing but the facts.” [P9]
\end{quote}

Specifically, as these participants envision the VUI adopting a friendly persona, their expectations encompass a desire for increased inclusion of social interaction elements, such as weather discussions [P14] and sharing of stories [P4, P5]. One participant [P6] proposed presenting users with a variety of VUI personas from which they could choose according to their personal preferences. 

\begin{figure}[H]
    \centering
    \includegraphics[width=\textwidth]{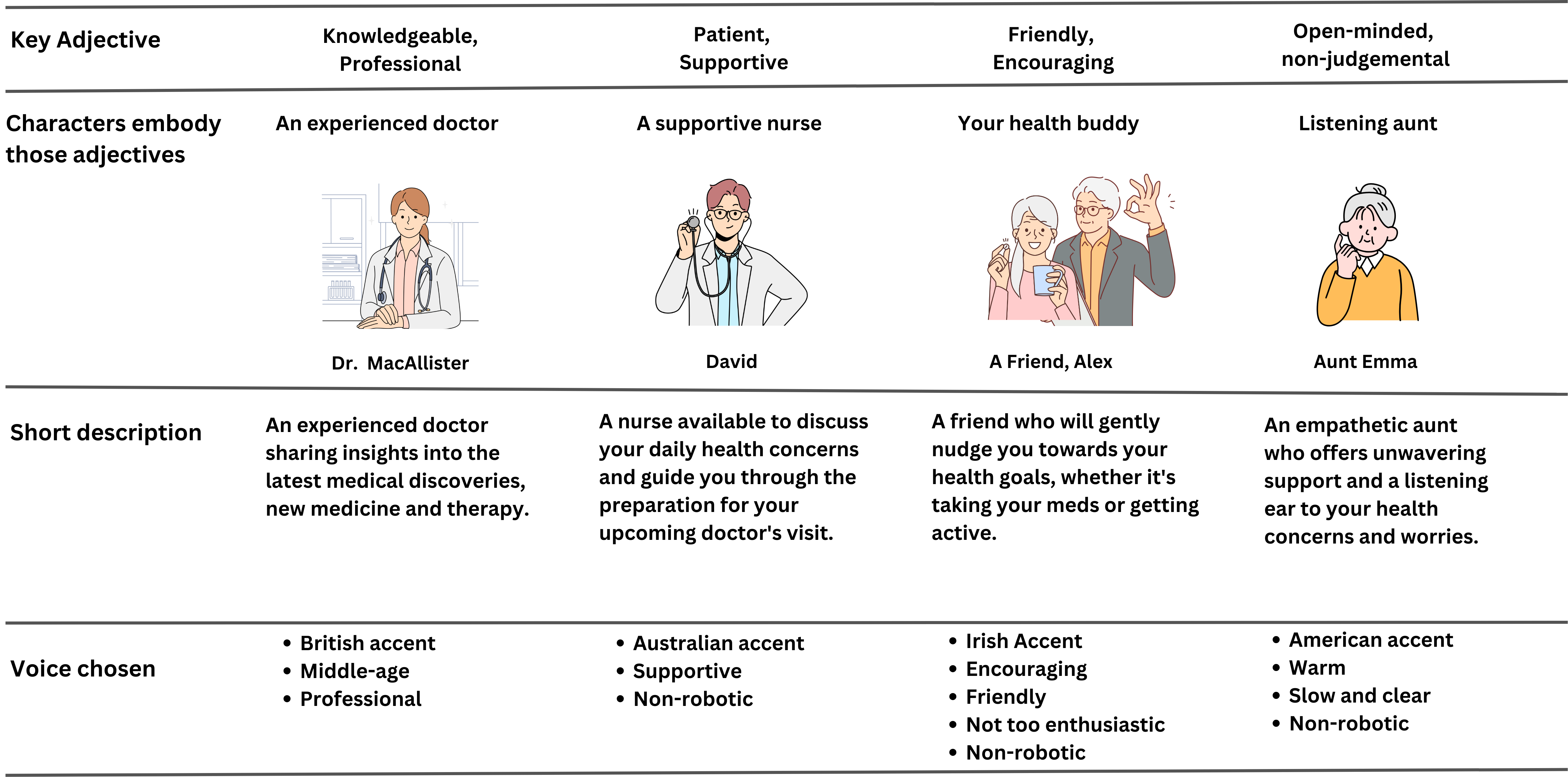}
    \caption{Personas developed during co-design sessions fused into four archetypes: an experienced doctor, a supportive nurse, a health buddy, and a listening aunt}
    \label{fig_key}
\end{figure}

The participants' preferences and insights are fused into 4 personas, as illustrated by Figure\ref{fig_key}

\subsubsection{Key Adjectives}

In addition to explicitly specifying personas, eight participants identified key adjectives that characterized the personality they designed for VUI, where friendliness, encouragement, open-mindedness, and reliability emerged as the most prominent ones.  Notably, when portraying the friendliness of VUI, not all participants intended to anthropomorphize it into a "friend." Instead, many described a quality more aligned with sociability and a welcoming attitude [P3, P10, P15]. It is also interesting that participants who assigned a friendly personality to VUI were also inclined to express a preference for encouragement [P10, P14] and open-mindedness [P14, P15] features in VUI. Their rationale suggests that combining these personality features will create a non-judgmental, patient, and caring impression of VUI. This becomes especially desirable in situations where older adults may miss information or encounter difficulties in managing health events. Furthermore, participants had two main motivations when designing a dependable personality trait for VUI: some believed it would enhance the credibility of VUI as an information source, while others saw it as a way to improve the perceived usefulness in monitoring users' health management.

\subsubsection{Voice}
While the voice of the VUI serves as an independent design element, participants have intentionally incorporated it into their persona design for the VUI. Regarding the voice aspect, the co-design process revealed some aesthetic and hedonistic considerations in user preferences for VUI besides the common concern about functionality. In both conditions, participants emphasized the design of the VUI's accent. Specifically, when designing the voice for VUI as an information provider, several participants chose accents such as Australian [P6], British [P2], and Irish [P8] for their own preferences.It is interesting to note that such non-functional, purely aesthetic, and hedonistic needs have been infrequently discussed by the participants in prior research. Furthermore, participants envisioned the VUI to offer customization options for accents to cater to a diverse user base. In particular, the participants expressed concerns about both the comprehensibility of VUI utterances to users [P8, P14, P18] and the VUI's ability to understand users [P1].

\begin{quote}
    “We reach out globally to people all over the world, so, there are different dialects (for VUI) to understand, as everybody likes to talk in different ways.” [P1]
\end{quote}

From the functional perspective, participants in both conditions designed a moderately calm yet non-robotic voice for VUI. After interacting with the learning probe, those who designed VUI favored a calm voice because it would sound affirming and credible when delivering information. In contrast, participants who designed VUI after interacting with the exercise probe aimed for a moderate level of calmness---avoiding extremes of hyperactivity while still maintaining a sense of engagement. Participants express that excessively calm and monotonous voices serve as a reminder to users that they are engaging with a machine, subsequently diminishing the perceived reliability of the VUI. Aligning with such concern, participants also explicitly pointed out that they would avoid the VUI having a robotic voice, which means a bland tone, choppy stops, and unnatural rhythms.

\section{Discussion}
\subsection{Co-designing with older adults: active involvement makes an impact}

In general, our study revealed that older adults become more effective in articulating their needs in the co-design process, consistent with prior design science research on collaborative innovation \cite{steen2013co}. We found that the co-design approach naturally guided the participants to adopt a design thinking mindset, organizing their ideations based on specific sets of design problems and solutions. Taking a conversational approach during the co-design process allowed for a narrative structure that more easily translated to corresponding design attributes as participants discussed and iterated their ideations with the researcher.  

We also observed that older adult participants can more effectively identify their Kansei needs in the co-design process and are more willing to share these preferences. An illustrative example is how participants designed the accents of VUI differently from previous studies. In prior research, including our own, participants’ consideration of accents in interaction with VUI primarily takes a usability stance, with rationales centered around improving the accessibility of VUI for a diverse global user base. However, in this study, participants naturally integrated their hedonistic preferences—such as Australian, Irish, and British accents—into the conceptualization of VUI accents. Their reasons spanned from desiring a trustworthy doctor with these accents to simply finding them charming. This change indicates that collaborating with older adults in designing has a significant impact on motivating them to express their wishes regarding their lifestyle aspirations, which may mitigate the concerns raised about biases in need assessment with older adults \cite{coleman2003design}. This is especially important given the persistent lack of awareness regarding the lifestyle aspirations of older adults has made sharing these preferences challenging for them, despite it being natural for younger adult users \cite{peine2014rise}.

Therefore, in alignment with previous studies that explored the possibility of co-design with older adults \cite{sakaguchi2021co, davidson2013participatory, hakobyan2013designing, sumner2021co}, we also found this approach to be valuable and meaningful. Nevertheless, we found that effectively taking the role of facilitator in co-design process, as urged in foundational collaborative design justice approaches \cite{costanza2020design}, required multiple strategies to both put co-designers in the designer's seat using care and humility strategies as in \cite{spiel2020thinking}, and also provide guidance through the short-turnover conversational design process our participants preferred. Guiding and probing questions, dictation, and sensory-rich opportunities were all found to be essential to avoid participants' design paralysis and to explore their hedonistic preferences and the design attributes needed to make these preferences a reality.

\subsection{Shifting Priorities with Social Model in Design}

Incorporating a social mental model that equally values non-functional and functional needs in VUI design reveals a notable shift in older adults' expressed priorities, particularly regarding socioemotional experiences. This aligns with the Socioemotional Selectivity Theory \cite{Löckenhoff_Carstensen_2004}, which posits that as we age, our priorities increasingly focus on socioemotional needs in regulating daily activities. Research has shown that older adults, compared to their younger counterparts, place greater value on socioemotional experiences as a means of protecting their well-being through engagement in selective environments and social interactions \cite{Charles_2010}. By encouraging older adults to consider non-functional needs during the early stages of design, we can ensure that VUIs are developed in a meaningful and relevant manner for this demographic.

For instance, while prior research acknowledged that participants view VUI 
accents as a critical design dimension, the focus was predominantly on functional aspects like clarity and understanding. However, when approached through a non-functional lens, considering social interaction, aesthetics, and psychological needs, participants actively proposed a desire for regional accents in VUIs, specifically mentioning British, Irish, and Australian accents as charming and engaging. This clearly highlights an important dimension that could significantly enhance user engagement and long-term interaction with VUIs. 

Moreover, when prompted to envision a VUI through a social mental model, older adult participants provided critical insights into how they would utilize the technology alongside their families and caregivers. They actively considered how the information provided by the VUI could benefit their entire support network in managing their health and well-being. For example, participants envisioned using the VUI to access information about health issues that could be shared and discussed with their children and grandchildren. They also considered the VUI's potential use within the specific layout and structure of their homes. These details and perspectives are crucial in understanding the multifaceted ways in which VUI technology can be integrated into older adults' lives. Such insights are often overlooked in studies that focus on older adults in isolation as a care-receiver.

Furthermore, this approach fostered greater openness and social sharing among older adults, particularly regarding non-functional needs and hedonic preferences. While previous research suggested a reluctance to share such information due to concerns about judgment and societal expectations \cite{desai2023using}, our findings reveal that framing the discussion within a social design context encourages older adults to openly express their aspirations and lifestyle preferences. This shift occurs as participants design for their peers, potentially reinforcing their own agency and sense of self through social sharing \cite{Lazar_Edasis_Piper_2017}. This rich understanding allows for the thoughtful integration of non-functional aspects into the design process from its inception, potentially leading to VUI designs that more comprehensively address the multifaceted needs of older adults. The willingness of participants to share these often-overlooked aspects of their preferences highlights the importance of creating supportive, judgment-free design contexts that empower older adults to fully express their desires and expectations for technology.

\subsection{Personas with stories}

Our findings reveal distinct preferences among older adults regarding VUI personas in different contexts, highlighting a potential pathway for establishing common ground between users and VUIs. When providing health information, participants predominantly preferred professional personas resembling experienced doctors, nurses, or researchers, while for exercise support, preferences were split between a health life coach and a warm, encouraging friend. This context-dependent preference suggests an opportunity for VUIs to adapt their personas dynamically, potentially enhancing the formation of a 'Theory of Mind' (ToM) \cite{Wang_Saha_Gregori_Joyner_Goel_2021} for both users and the VUI system. For instance, the VUI could present a doctor persona when the user is seeking information about disease symptoms, transitioning to a friendlier, more empathetic persona when the user expresses worry about these symptoms. This adaptive approach could foster a sense of familiarity and understanding, potentially leading to more effective and engaging interactions.

An intriguing observation is how older adults perceive the relationship between personas and the VUI itself. Participants' rationales suggest they did not expect the VUI to maintain a consistent persona across various situations, indicating a sophisticated understanding of the VUI's potential to embody multiple roles. This flexibility in perception aligns with the concept of a malleable ToM, where users can adjust their expectations and interpretations of the VUI's intentions and knowledge based on the context of the interaction. The inconsistency in persona attributes across scenarios, repeatedly observed throughout the co-design process, suggests that older adults may envision the VUI as a collection of different personas, each suited to specific interaction contexts. Our findings align with prior research by Desai \& Twidale \cite{Desai_Twidale_2023} suggesting that users develop more sophisticated mental models of human-VUI interactions than current design processes typically account for, underscoring the need for a more nuanced design approach that can accommodate and leverage this complexity to create VUIs that more effectively meet users' dynamic expectations and needs.

Furthermore, older adults demonstrated a strong inclination to enrich the background stories of VUI personas, even when these details lacked functional significance. This preference for elaborate backstories, such as a doctor persona with 30 years of experience in heart diseases and knowledge of recent Harvard Medical magazine breakthroughs, offers insights into how users might form a more complex ToM for VUIs. By providing these detailed narratives, users may be attempting to create a more comprehensive mental model of the VUI's knowledge and experience, potentially leading to increased trust and engagement. This enrichment of personas could serve as a bridge for finding common ground, allowing users to relate to the VUI on a more personal level. From the VUI's perspective, incorporating these rich backstories into its adaptive persona system could enhance its ability to present as a knowledgeable, experienced entity, potentially increasing user confidence in its capabilities and advice. 

\subsection{Limitations and Future Work}
Our study offers valuable insights into VUI design for older adults, but we acknowledge limitations that present opportunities for future research. Our participant pool, primarily older adults from a Midwest college town accessible via email, may not fully represent this diverse demographic, necessitating future studies with broader, more diverse groups to enhance generalizability. While our one-to-one co-design process was insightful, future work could complement this with group-based sessions to capture collective insights. Additionally, longer-term interaction with VUIs in naturalistic settings would yield more ecologically valid outcomes than the brief familiarization period provided. Our immediate next steps involve exploring participants' mental models of VUI personas, particularly investigating how narratives shape these models and influence user engagement. The ultimate goal is to develop evidence-based guidelines for VUI design that enhance older adults' well-being, health, and eudaimonia. Future research should examine how preferences and mental models translate into actual usage behaviors and health outcomes in real-world scenarios, contributing to more inclusive and effective VUI technologies for the aging population. By addressing these limitations and pursuing these research directions, we aim to significantly advance the field of VUI design for older adults, creating technologies that are functional but also personally meaningful and easily integrated into their daily lives.

\section{Conclusion}
In conclusion, this study engaged older adults in co-design activities, prompting them to conceptualize their ideal VUI in health-related contexts. The outcomes significantly deepened our understanding of the sensory, emotional, and aesthetic preferences of older adults for VUI, particularly in terms of personality, voice, and narratives, among other dimensions. Notably, the co-design process played a pivotal role in eliciting and articulating the Kansei needs of older adults for VUI. Beyond mere functionality considerations, our findings illustrate the significance of aligning VUI design with the lifestyle aspirations of older adults.


\bibliographystyle{ACM-Reference-Format}
\bibliography{sample-base}

\end{document}